\documentclass[prl,twocolumn,showpacs,amsmath,amssymb,superscriptaddress]{revtex4}

\usepackage{graphicx}
\usepackage{dcolumn}
\usepackage{bm}

\begin{document}

\title{Valley Polarization in Si(100) at Zero Magnetic Field}

\author{K. Takashina}
 \affiliation{NTT Basic Research Laboratories,
NTT Corporation, Atsugi-shi, Kanagawa 243-0198, Japan}

\author{Y. Ono}
 \affiliation{NTT Basic Research Laboratories,
NTT Corporation, Atsugi-shi, Kanagawa 243-0198, Japan}

\author{A. Fujiwara}
 \affiliation{NTT Basic Research Laboratories,
NTT Corporation, Atsugi-shi, Kanagawa 243-0198, Japan}

\author{Y. Takahashi}
\altaffiliation[Present Address: ]{Graduate School of Information
Science and Technology, Hokkaido University, Kita 14, Nishi 9,
Kita-ku, Sapporo, 060-0814 Japan}
 \affiliation{NTT Basic Research Laboratories,
NTT Corporation, Atsugi-shi, Kanagawa 243-0198, Japan}

\author{Y. Hirayama}
 \affiliation{NTT Basic Research Laboratories,
NTT Corporation, Atsugi-shi, Kanagawa 243-0198, Japan}
 \affiliation{ SORST JST, Kawaguchi, Saitama 331-0012, Japan}

\date{\today}

\begin{abstract}
The valley splitting, which lifts the degeneracy of the lowest two
valley states in a SiO$_2$/Si(100)/SiO$_2$ quantum well is examined
through transport measurements. We demonstrate that the valley
splitting can be observed directly as a step in the conductance
defining a boundary between valley-unpolarized and polarized
regions. This persists to well above liquid helium temperature and
shows no dependence on  magnetic field, indicating that
single-particle valley splitting and valley-polarization exist in
(100) silicon even at zero magnetic field.

\end{abstract}

\maketitle


Beyond its immense technological importance, silicon continues to
command considerable physical interest. This, on one hand, is
driven by relentless technological advances producing
ever-higher-quality and versatile structures such as Silicon on
Insulator (SOI) allowing unexplored physical regimes to be
examined, and complementary processing techniques facilitating new
experimental schemes\cite{SOIexample}. On the other hand, silicon
possesses a number of unique properties, a very fundamental one of
which is its electronic band structure.

The conduction band dispersion relation of bulk silicon has 6
equivalent minima lying along three mutually orthogonal axes of the
Brilloin Zone. With two-dimensional confinement parallel to one of
these axes, this six-fold degeneracy is lifted such that the two
minima lying along the confinement axis become the lowest lying
states due to anisotropy of effective
mass\cite{AndoFowlerSternReview}. Electrons in 2-dimensional
structures such as Si(100) MOSFETs (Metal Oxide Semiconductor Field
Effect Transistors) occupy these remaining two valleys such that
each electron has both valley and spin degrees of freedom. In recent
years, this remaining valley degeneracy and valley splitting which
lifts this degeneracy have experienced a revival of interest, partly
because valleys offer a relatively unexplored degree of freedom in
relation to metal-insulator transitions\cite{MetInsValleyReview} and
other many-body
physics\cite{ZeitlerLaiPRL,PudalovCondMat,KhrapaiPRB,ShkolnikovPRL,SiGeEVR},
but also due to new device
possibilities\cite{etoPRB,BoykinAPL,AhnCondMat} and as a source of
decoherence in spin-based quantum
computation\cite{BoykinAPL,DasSarma,SiGeEVR}.

Although effects indicating interactions between valleys have been
known for a long time\cite{FowlerFangHowardStilesPRL}, they had been
limited to structures involving vicinal planes which break the
in-plane symmetry, or in (100) structures, special experimental
conditions of low temperature and magnetic
field\cite{AndoFowlerSternReview}. Despite the long history, a
consensus for the nature of this valley splitting has yet to be
reached. In particular, some
studies\cite{ShkolnikovPRL,KhrapaiPRB,PudalovJETP,SiGeEVR} have
pointed to a strong, almost linear, dependence on magnetic field,
suggesting a many-body origin as opposed to a single-particle one,
and its existence without magnetic field in Si(100) has not been
experimentally confirmed.

In this Letter, we show the first direct and vivid evidence for
valley splitting of Si(100) in the absence of magnetic field. Recent
high magnetic field experiments have found that the Si/SiO$_2$
interface in SIMOX (Separation by IMplantation of OXygen (a type of
SOI)) structures offer values of
valley-splitting\cite{ValleySIMOXOuisse,ValleySIMOX} of at least a
couple of meV, large compared with values found in conventional
Si-MOS structures under equivalent conditions. Here, we show that
the valley splitting in these structures can be increased to 10's of
meV, and that this large valley splitting leads to a step feature in
the conductance defining a boundary between valley-unpolarized and
valley-polarized regions. This allows us to verify the existence of
valley-splitting in the absence of magnetic field. Further, we find
its effects to be important well above liquid helium temperatures
and that the valley splitting can be approximated by a linear
dependence on electrical bias even at unprecedentedly large values
found here. These findings strongly suggest its single-particle
nature, and demonstrate its potential for extensive band structure
control and for examining new physical regimes where valley
splitting is comparable to, or greater than, otherwise dominant
energy scales.


\begin{figure}[b]
\includegraphics[width=1\linewidth]{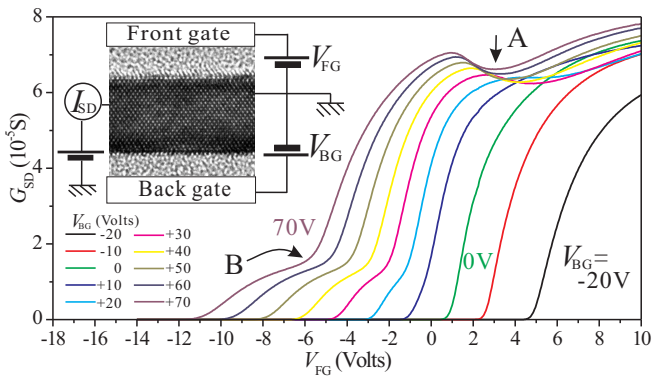}
\caption{\label{Fig1}(\textit{Color online}) The source-drain
conductance $G_{\mathrm{SD}}$ at 4.2K as a function of
$V_{\mathrm{FG}}$ at different values of $V_{\mathrm{BG}}$. A and B
mark features due to second spatial subband and upper valley subband
occupation respectively. The inset shows the experimental setup with
a TEM image of a SIMOX structure.}
\end{figure}

The samples consist of SiO$_2$/Si(100)/SiO$_2$ quantum well
transistors [Fig.1 (\textit{inset})] fabricated on SIMOX substrates
where the buried oxide is formed through ion-implantation followed
by high temperature annealing\cite{ValleySIMOX}. Data presented here
come from a nominally 8nm wide quantum well clad between thermal and
buried oxide layers 75nm and 380nm thick respectively, with large
device dimensions of 30$\mu$m channel-width and 208$\mu$m between
source and drain. The peak Hall mobility in these structures is
$0.8$m$^2$/Vs. Front and back-gate capacitances were found to be
$C_\mathrm{F}=485\mu$Fm$^{-2}$ and $C_\mathrm{B}=92\mu$Fm$^{-2}$
respectively agreeing with sample dimensions.

The source-drain current $I_{\mathrm{SD}}$ was measured as a
function of front gate voltage $V_{\mathrm{FG}}$ at various values
of back gate voltage $V_{\mathrm{BG}}$ [Fig.1] (at a constant source
voltage (3mV) and constant temperature (4.2K),
$G_\mathrm{SD}=I_{\mathrm{SD}}/3$mV). With $V_{\mathrm{BG}}$ set at
zero volts, $G_{\mathrm{SD}}$ shows behavior characteristic of a
standard silicon MOSFET at low temperature. At first, carriers
induced by $V_{\mathrm{FG}}$ are localized and do not contribute to
conduction. With the electron number density $n$ increased, they
begin to conduct and $G_\mathrm{SD}$ shows a sharp increase. When
negative $V_{\mathrm{BG}}$ is applied, the overall trace shifts to
higher $V_{\mathrm{FG}}$ since the carrier concentration $n$ also
depends on $V_\mathrm{BG}$ according to
$n=C_\mathrm{F}(V_\mathrm{FG}-V^\mathrm{Th}_\mathrm{F})+C_\mathrm{B}(V_\mathrm{BG}-V^\mathrm{Th}_\mathrm{B})$
where $V^\mathrm{Th}_\mathrm{F}$ and $V^\mathrm{Th}_\mathrm{B}$ are
constant offsets. Also, since the electric field increases, a
shallower slope in $G_\mathrm{SD}(V_\mathrm{FG})$ results due to
increased interface roughness scattering. Behavior at
$V_\mathrm{BG}\leq0$ is virtually identical to conventional MOSFETs
with substrate bias\cite{AndoFowlerSternReview}.

With positive $V_{\mathrm{BG}}$, $G_\mathrm{SD}$ traces shift to
lower $V_\mathrm{FG}$ and a minimum develops [A in Fig.1] due to the
second subband, or alternatively worded, electron channels being
present at both front and back interfaces \cite{selfconc}. These two
layers couple to form bonding and antibonding states, hereon
referred to as ``spatial subbands", and the extent to which these
should be described as two ``electric" subbands or two spatially
separated layers depends on the potential bias of the quantum well
and self-consistent effects\cite{ValleySIMOX,selfconc}. While
interlayer coupling changes the wavefunction thereby affecting
mobility \cite{PalevskiResistanceResonance,Prunnila}, the
accessibility of the second subband allow further scattering
processes\cite{StormerInterSubbandScattering,locSpinScat} which
reduces the conductance. These inter-spatial-subband effects, which
have already been examined in other systems\cite{Prunnila}, are
expected to be involved in producing the minimum here.

Positive $V_\mathrm{BG}$ also shifts the wavefunction of the lower
spatial subband closer to the buried-oxide interface with greater
valley splitting\cite{ValleySIMOX}. With large positive values of
$V_{\mathrm{BG}}$, the data reveal another feature [B in Fig.1]. We
interpret this also to be caused by intersubband effects, but rather
than spatial subbands, we argue that this feature arises due to
valley splitting leading to two valley subbands. At low
$V_\mathrm{FG}$, the conduction is solely due to the lower valley
subband and the step feature represents the onset of occupation of
the upper valley\cite{ShkolnikovPiezoZeroB}.

\begin{figure}[t]
\includegraphics[width=1\linewidth]{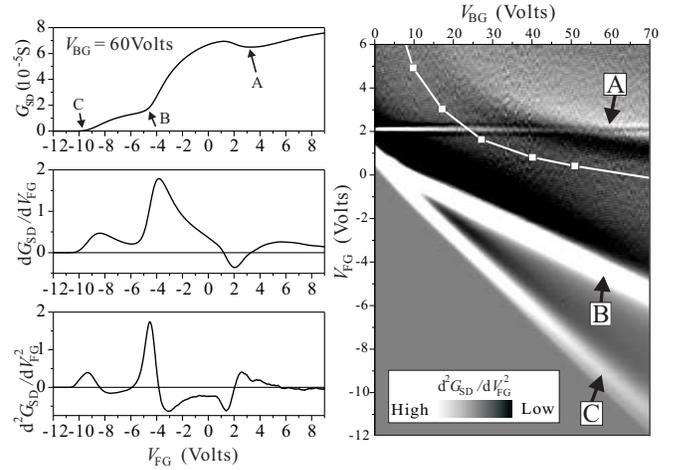}
\caption{\label{Fig2} \textit{Left}: The conductance at
$V_{\mathrm{BG}}=60$V, 4.2K. The lower two graphs show
1$^\mathrm{st}$ and 2$^\mathrm{nd}$ derivatives. The feature marked
A is due to the second spatial subband. B marks the feature due to
the upper valley while C marks the onset of conduction.
\textit{Right}: Gray-scale plot of the 2$^\mathrm{nd}$ derivative
showing the evolution of A, B and C with
($V_{\mathrm{BG}},V_{\mathrm{FG}}$). White squares show calculated
onsets of second spatial-subband occupation \cite{selfconc}. The
abrupt horizontal line at $V_\mathrm{FG}=2$V is an experimental
artifact. }
\end{figure}


In order to examine the evolution of these features with
$V_{\mathrm{BG}}$, we doubly differentiate the
$G_\mathrm{SD}(V_{\mathrm{FG}})$ traces to accentuate the features
[Fig.2, \textit{left}] and construct gray-scale plots as a function
of $(V_{\mathrm{BG}},V_{\mathrm{FG}})$ [Fig.2, \textit{right}]. The
initial onset of conduction and the subsequent step both appear as
peaks in the double-differential, and as white regions on the
gray-scale plot. While peak C at low $V_\mathrm{FG}$ marks the onset
of conduction, peak B defines the boundary between valley polarized
and unpolarized regions. It is readily seen that with increased
$V_\mathrm{BG}$, these peaks separate, and appear to continue
separating in an approximately linear manner. The valley-polarized
region grows in range as $V_\mathrm{BG}$ is increased indicating a
linear increase in valley splitting $\Delta$. This linear increase
of $\Delta$ with $V_\mathrm{BG}$ and hence electric field is
qualitatively consistent with theoretical
predictions\cite{AndoFowlerSternReview,BoykinAPL} and previous
magneto-transport estimates of the valley
splitting\cite{NichBGValley}.


To confirm that the structure arises due to valley-splitting, we
performed measurements in magnetic field. Features similar to A,B
and C at zero magnetic field can be seen at approximately the same
$(V_\mathrm{BG},V_\mathrm{FG})$ at $B=5.5$T (also marked A,B and C
in Fig. 3(a)), demarcating regions showing different patterns of
Shubnikov de Haas (SdH) oscillations. With positive
$V_{\mathrm{FG}}$ and small $V_{\mathrm{BG}}$ (top-left of the
graph) where the valley splitting $\Delta$ is known to be small, SdH
oscillations can be seen with a period determined by the total
filling factor $\nu^{\mathrm{tot}} = nh/eB$. Minima are seen
whenever $\nu^{\mathrm{tot}}$ is an integer multiple of 4, since 4
is the combined degeneracy of valley and
spin\cite{AndoFowlerSternReview}. In contrast, in the region lying
between features B and C, the oscillations occur with half the
periodicity of $\Delta\nu^{\mathrm{tot}}=2$, showing that the only
remaining degeneracy here is due to spin and the system is
completely valley-polarized where the electrons only occupy the
lower valley subband. In the region above B, two sets of
oscillations can be seen. One set describes lines rather parallel to
B, verifying that B represents the onset of the upper valley subband
\cite{secondSubbandValley}.


\begin{figure}[t]
\includegraphics[width=1\linewidth]{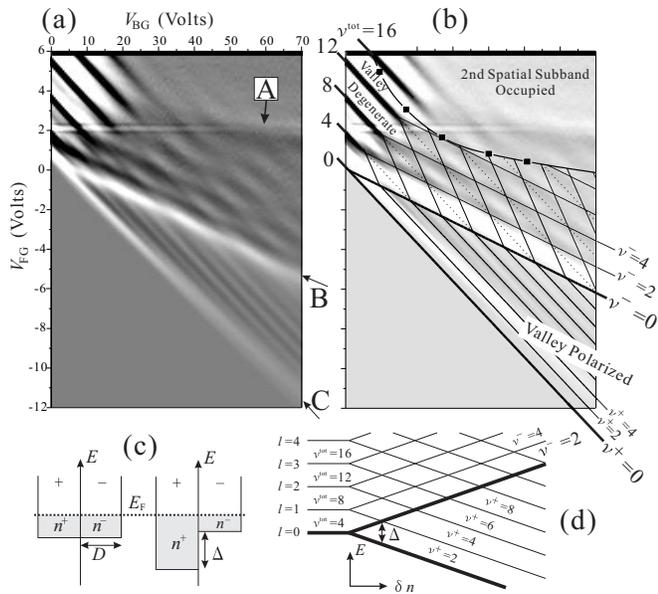}
\caption{\label{Fig3} (a) d$^2G_{\mathrm{SD}}/$d$V_{\mathrm{FG}}^2$
at 5.5T, 4.2K. (b) Same data as (a) but with different contrast.
Solid squares joined by curves are estimated onsets of
2$^\mathrm{nd}$ spatial subband occupation
\cite{selfconc,secondSubbandValley}. Solid straight lines show
calculated lines of even $\nu^+$ and $\nu^-$ while dotted lines
represent $\nu^{\mathrm{tot}}=4i$ where $i$ is an integer \cite{4i}.
The region beneath $\nu^-=0$ is completely valley-polarized. (c)
Density of states at zero magnetic field with and without valley
splitting $\Delta$. (d) Resultant Landau levels. Thicker lines also
correspond to valley subband edges.}
\end{figure}

To make estimates of $\Delta$, we use a simple phenomenological
model. If each valley has identical and constant in-plane effective
mass $m^*$, the density of states for each subband is
$D=m^*m_\mathrm{0}/\pi\hbar^2$. When both valleys have non-zero
occupation, the number of electrons in each valley $n^\pm$ is then
given by $n^\pm = \frac{1}{2}(n\pm D\Delta)$ in which superscripts +
and - indicate lower and upper valley states respectively
[Fig.3(c)]. Since the directions along which the oscillations are
occurring appear rather linear in
$(V_{\mathrm{FG}},V_{\mathrm{BG}})$, we suppose a simple empirical
approximation: $\Delta = \alpha \delta n$ when $\delta n >0$ where
$\delta n = n_{\mathrm{B}}-n_{\mathrm{F}}$, $n_{\mathrm{B}}$ and
$n_{\mathrm{F}}$ being electron densities contributed by the back
and front gates, respectively. (We set $\Delta=0$ for $\delta n<0$
since it is negligibly small.) It then follows that
\begin{equation}\label{eqn:Fit}
    n^\pm = \frac{1}{2}[(1\mp D \alpha)n_{\mathrm{F}}+(1\pm D \alpha)n_{\mathrm{B}}]
\end{equation}
which can be used to calculate the filling factors of the lower and
upper valleys $\nu^+$ and $\nu^-$ respectively under magnetic field
[Fig.3(d)]. Solid lines in Fig.3 (b) show regions where $\nu^+$ and
$\nu^-$ are even, where the fitting parameter has been set to
$\alpha=0.46$meV$/10^{15}$m$^{-2}$ chosen to fit the minimum along
$\nu^-=2$. Although this value of $\alpha$ works reasonably well
between $\nu^-=0$ and $\nu^-=4$, smaller values of around
$\alpha=0.40$meV$/10^{15}$m$^{-2}$ work better for higher filling
factors between $\nu^-=8$ and 10. This is likely to be due to the
larger concentration leading to more screening, reducing the effect
of the external bias. In turn, it also suggests that greater values
are present at lower concentrations.

We have also performed measurements where the magnetic field was
swept keeping $V_\mathrm{BG}$ constant at 60V which have shown that
the SdH oscillations can be fitted to similar accuracy as the
constant field data [Fig.3] without any magnetic field dependence of
$\Delta$. This confirms that the valley splitting here is
independent of magnetic field, corroborating the correspondence
between the data at zero field [Fig.2] and higher field [Fig.3].

\begin{figure}[b]
\includegraphics[width=1\linewidth]{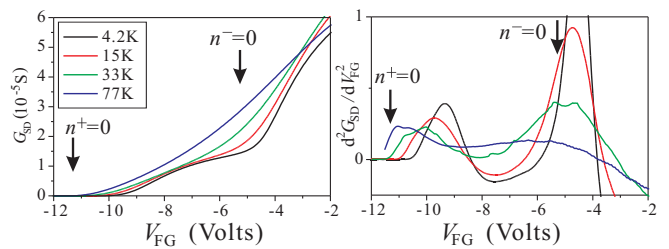}
\caption{\label{Fig4} (\textit{Color online}) Temperature dependence
at $B=0$T, $V_\mathrm{BG}=60$V. \textit{Left}: $G_\mathrm{SD}$.
 \textit{Right}:
d$^2G_{\mathrm{SD}}/$d$V_{\mathrm{FG}}^2$. Arrows mark valley
subband edges extracted using $\alpha=0.46$meV$/10^{15}$m$^{-2}$.}
\end{figure}

Fig.4 shows $G_\mathrm{SD}$ and its second derivative taken at zero
magnetic field and $V_\mathrm{BG}=60$V where we estimate $\Delta
=23$meV ($\Delta/k_B=270$K) at the onset of upper valley occupation.
The valley subband edges expected from the fit are also marked,
suggesting that the actual bottom of the upper valley subband
($n^-=0$) occurs at a lower value of $V_\mathrm{FG}$ than the center
of the peak in d$^2G_\mathrm{SD}/$d$V_\mathrm{FG}^2$. This is
consistent with the interpretation that the step feature in
$G_\mathrm{SD}$ arises as a consequence of suppression in mobility
when the upper valley becomes energetically accessible. The
step-like increase occurs when this suppression becomes weaker,
after the onset of occupation, as the Fermi energy moves further
above the subband edge. However, further work to establish the roles
played by inter-valley interactions and the differences in mobility
of the two individual valley subbands are required before a
quantitative understanding of $G_\mathrm{SD}$ can be reached.

Fig. 4 also shows that the conduction band edge expected from the
fit lies at lower $V_\mathrm{FG}$ than the onset of conduction. In
Fig. 3, it is also seen, that although lines of constant $\nu^+$
describe straight diagonal lines, the peak associated with the onset
of conduction (C) curves toward higher $V_\mathrm{FG}$ with
increased $V_\mathrm{BG}$. This can be attributed to increase in the
number of localized states with $V_\mathrm{BG}$ as the disorder is
increased, due in turn to the electrons being pressed against the
Si-SiO$_2$ interface. With increased temperature, the peak in the
second differential shifts toward the predicted conduction band
edge.

At higher temperatures, the feature associated with the upper
valley can still be discerned directly in $G_\mathrm{SD}$ at 33K,
but it is not so clear at 77K. However, the double differential
does show the second peak even at 77K. At such a temperature and
above, the conduction globally becomes strongly influenced by
phonon scattering and is no longer a good probe for detecting
subband edges.

We now discuss our findings in relation to recent experiments in
other systems. Studies on conventional Si MOS\cite{KhrapaiPRB} and
Si/SiGe\cite{SiGeEVR} have both shown very strong dependence of
$\Delta$ on magnetic field. In Si/SiGe, the absolute values of
$\Delta$ are extremely small (10's of $\mu$eV) suggesting that their
phenomena belong to a completely different regime at very much
smaller energy scales. In conventional Si MOS structures, $\Delta$
has relatively larger values between 0.2 and 2meV, but these are
still smaller than values addressed in this work. Furthermore, the
reported field dependence comes from measurements of energy gaps in
the quantized Hall regime\cite{KhrapaiPRB,PudalovJETP} and their
results are likely to be heavily influenced by many-body effects
dominant under their conditions.

The results here suggest, therefore, that there exists a
\textit{bare} valley splitting arising from single particle
interactions \cite{BoykinAPL}, independent of magnetic field. When
it is of the order of a few meV and greater, it dwarfs many body
contributions and there is very little magnetic field dependence.
When the bare splitting is small, on the other hand, interaction
effects become dominant and heavily influences the energy-gaps and
its dependence on magnetic field.

There is also another important conclusion we can draw from the
present results. In our previous experiments, we could only
conclude that the valley splitting at the Si-buried oxide
interface was much larger compared with the standard thermal-oxide
interface at the front showing usual small
values\cite{ValleySIMOX}. The linear dependence on electrical bias
we find here, demonstrates that this rule is applicable even when
the absolute value of $\Delta$ becomes orders of magnitude greater
than those normally observed. What changes from one type of
interface to another is the prefactor $\alpha$.

This work is partially supported by JSPS KAKENHI (16206003) and
(16206038).


\begin{thebibliography}{0}
\bibitem{selfconc} We have performed self-consistent calculations
neglecting valley-splitting and manybody effects to estimate
($V_\mathrm{FG},V_\mathrm{BG}$) at the onset of 2$^\mathrm{nd}$
subband occupation for an 8nm wide well. These are presented in
Figs. 2 and 3(b) and agrees qualitatively with the behavior of
feature A.
\bibitem{secondSubbandValley} We observe no valley splitting feature
like B for the upper spatial subband. This is because the
wavefunction of the second spatial subband is never large close to
the buried oxide interface in the accesible gate-voltage range.
\bibitem{4i}We note that deepest SdH minima are seen when conditions
$\nu^\mathrm{tot}=4i$ and $\nu^-=2i$ are both met while those where
$\nu^-=2i$, $\nu^+=2i$ but $\nu^\mathrm{tot}\neq 4i$ are not as
pronounced. We speculate that this is due to many-body effects which
occur at Landau level coincidences \cite{JunwirthMacDonald}.
\end{thebibliography}
\end{document}